\documentclass[12pt]{article}
\pdfoutput=1
\usepackage[english]{babel}
\usepackage[latin1]{inputenc}
\usepackage[normalem]{ulem}
\usepackage{cite} 
\usepackage{pdfcolmk}
\usepackage[intlimits]{amsmath}
\usepackage{amssymb}
\usepackage{graphicx}
\usepackage{color}

\usepackage[text={16.4cm,23.7cm}]{geometry}

\newcommand{\GeV}{\,\mbox{GeV}}
\newcommand{\TeV}{\,\mbox{TeV}}

\newcommand{\cm}{\,\mbox{cm}}
\newcommand{\m}{\,\mbox{m}}
\newcommand{\s}{\,\mbox{s}}
\newcommand{\sr}{\,\mbox{sr}}
\newcommand{\kpc}{\,\mbox{kpc}}
\newcommand{\Myr}{\,\mbox{Myr}}
\newcommand{\MV}{\,\mbox{MV}}
\newcommand{\GV}{\,\mbox{GV}}

\begin{document}

\title{\vspace{-2cm} 
{\normalsize
\flushright TUM-HEP 905/13\\}
\vspace{0.6cm} 
\bf Dark matter annihilations and decays after the AMS-02 positron measurements \\[8mm]}
\author{Alejandro Ibarra$^1$, Anna S. Lamperstorfer$^1$, Joseph Silk$^{2,3,4}$ \\[2mm]
{\normalsize\it $^1$Physik-Department T30d, Technische Universit\"at M\"unchen,}\\[-0.05cm]
{\normalsize \it James-Franck-Stra\ss{}e, 85748 Garching, Germany}\\[-0.05cm]
{\normalsize\it $^2$Institut d'Astrophysique, UMR 7095 CNRS, Universit\'{e} Pierre et Marie Curie, }\\[-0.05cm]
{\it\normalsize 98bis Blvd Arago, 75014 Paris, France}\\[-0.05cm]
{\normalsize\it $^3$Department of Physics and Astronomy, The Johns Hopkins University,}\\[-0.05cm]
{\it\normalsize Homewood Campus, Baltimore MD 21218, USA}\\[-0.05cm]
{\normalsize\it $^4$Beecroft Institute of Particle Astrophysics and Cosmology,}\\[-0.05cm]
{\it\normalsize  Department of Physics, University of Oxford, Oxford OX1 3RH, UK}
}
\maketitle

\begin{abstract}
The AMS-02 Collaboration has recently presented high-quality measurements  of the cosmic electron and positron fluxes as well as the positron fraction. We use the measurements of the positron flux to derive, for the first time, limits on the dark matter annihilation cross section and lifetime for various final states. Working under the well-motivated assumption that a background positron flux exists from spallations of cosmic rays with the interstellar medium and from astrophysical sources, we find strong limits on the dark matter properties which are competitive, although slightly weaker, than those derived from the positron fraction. Specifically, for dark matter particles annihilating only into $e^+ e^-$ or into $\mu^+\mu^-$, our limits on the annihilation cross section are  stronger than the thermal value when the dark matter mass is smaller than 100 GeV or  60 GeV, respectively.
\end{abstract}

\section{Introduction}
\label{sec:introduction}

In a large class of dark matter (DM) scenarios, the annihilations or decays of the dark matter particles produce a positron flux, accompanied by an electron flux, thus opening the possibility for the indirect detection of dark matter in our Galaxy through  observation of an excess in the cosmic positron flux or positron fraction relative to the expected astrophysical backgrounds. In the last few years, data of extraordinary quality on the positron flux or fraction has been collected by several experiments, notably  CAPRICE~\cite{Boezio:2000zz}, HEAT~\cite{DuVernois:2001bb}, AMS-01~\cite{Aguilar:2007yf}, PAMELA~\cite{Adriani:2008zr,Adriani:2013uda} and AMS-02~\cite{Aguilar:2013qda,Schael}. The most striking result is the observation of a rise in the positron fraction starting at $\sim 10\GeV$ and extending at least to 350 GeV, as recently confirmed by the AMS-02 Collaboration~\cite{Aguilar:2013qda}. While dark matter annihilations or decays provide a possible explanation for this observation, the cosmic positron backgrounds are still poorly understood and it is not possible at the moment to make a definite statement about the origin of this excess.  

From a conservative point of view, the measurements of the positron flux or fraction allow one to set limits on the dark matter annihilation cross section and decay width in a given dark matter scenario. Many studies have derived limits on the dark matter parameters employing measurements of the positron fraction under the assumption that the positron background vanishes or that the only source of background is  secondary production due to spallations of cosmic rays in the interstellar medium (see {\it e.g.} Refs. \cite{Cirelli:2008pk,Ibarra:2008jk,Meade:2009iu,Kopp:2013eka,Feng:2013zca,Yuan:2013eja,Cholis:2013psa,Jin:2013nta}). This approach was recently improved in Ref. \cite{Bergstrom:2013jra}, where  a search was undertaken  in the positron fraction for  spectral features which are predicted to arise in certain final states of dark matter annihilations. We will consider in this paper the limits that can be obtained from the positron flux and which, in contrast to the positron fraction, do not require knowledge of the electron flux and are therefore cleaner from the theoretical point of view. These limits are complementary (and as we will show, competitive) to those derived from the positron fraction. We will also include in our calculation a plausible background model to take into account a possible primary component of astrophysical origin in the positron flux, which is likely to exist and which presumably provides the additional source of positrons when the dark matter annihilations or decays cannot totally explain the rise in the positron fraction (for example, when the dark matter mass is smaller than 300 GeV or when the annihilation cross section or decay rate is too small). Besides, most analyses neglect the positron data at low energies  (see however  recent analyses \cite{Hooper:2012gq,Bergstrom:2013jra} where the low energy data of the positron fraction was used). Nevertheless, the exquisite measurements provided by experiments, together with the excellent agreement of the observations with the expectations from non-exotic physics which leave little room for an exotic component in the positron fraction, allow us to set very stringent bounds on the dark matter properties.

The paper is organized as follows: in section \ref{sec:fluxes}, we review the methods to calculate  positron fluxes at the Earth from dark matter annihilations or decays and we present an ansatz for the background electron and positron fluxes. In section \ref{sec:limits} we present our calculation of the limits on the annihilation cross section and lifetime and we apply it to various possible final states. Lastly, in section \ref{sec:conclusions} we present our conclusions.

\section{Primary and secondary positron fluxes at the Earth}
\label{sec:fluxes}

The annihilations or decays of dark matter particles in the Milky Way dark matter halo at the position $\vec r$ with respect to the Galactic center produce a primary positron flux  with a rate per unit energy and unit volume which is given by (for reviews, see Refs. \cite{Salati:2010rc,Ibarra:2013cra}):
\begin{align}
Q_{e^+}(E,\vec r)=
\begin{cases} \displaystyle{ \frac{1}{2}\frac{\rho_{\rm DM}^2(\vec r)}{m^2_{\rm DM}}
\sum_f \langle \sigma v\rangle_f \frac{dN^f_{e^+}}{dE}} & {\rm (annihilations)}\;, \\
\displaystyle{\frac{\rho_\text{DM}(\vec{r})}{m_{\rm DM}}\sum_f \Gamma_f\frac{dN^f_{e^+}}{dE}} & {\rm (decays)}\;,
\end{cases}
\label{eq:source}
\end{align}
where $\langle \sigma v\rangle_f$ denotes the velocity-weighted annihilation cross section, $\Gamma_f$ the decay rate and $dN^f_{e^+}/dE$ the energy spectrum of positrons produced in the annihilation or decay channel $f$, which we simulate using the event generator PYTHIA 8.1 \cite{Sjostrand:2007gs}. Besides, $\rho_\text{DM}(\vec{r})$ denotes the density of dark matter particles in the Milky Way halo, which we assume to be spherically symmetric and with a radial distribution described by the Einasto profile~\cite{Navarro:2003ew,Graham:2005xx,Navarro:2008kc},
\begin{equation}
  \rho_\text{DM}(r)=\rho_0 \exp\left[-\frac{2}{\alpha}
\left(\frac{r}{r_s}\right)^\alpha\right]\;,
\end{equation}
with $\alpha=0.17$ and $r_s=28 ~\rm{kpc}$. The overall normalization factor $\rho_0$ will be chosen to reproduce the local dark matter density $\rho_\odot= 0.39~\text{GeV}/\text{cm}^3$~\cite{Catena:2009mf,Weber:2009pt,Salucci:2010qr,Pato:2010yq,Iocco:2011jz} with $r_\odot=8.5$ kpc.

The propagation of positrons in the Milky Way is commonly described by a stationary two-zone diffusion model with cylindrical boundary conditions~\cite{Berezinskii}. Under this approximation, the number density of particles per unit energy, $f_{e^+}(E,\vec{r},t)$, satisfies the following transport equation:
\begin{equation}
0=\frac{\partial f_{e^+}}{\partial t}=
\nabla \cdot [K(E,\vec{r})\nabla f_{e^+}] +
\frac{\partial}{\partial E} [b(E,\vec{r}) f_{e^+}]+Q(E,\vec{r})\;,
\label{transport}
\end{equation}
with boundary conditions such that the solution  $f_{e^+}(E,\vec{r},t)$ vanishes at the boundary of the diffusion zone, which is approximated by a cylinder with half-height $L = 1-15~\rm{kpc}$ and radius $R = 20 ~\rm{kpc}$. In this equation, $K(E,\vec r)$ is the diffusion coefficient which we assume constant throughout the diffusion zone and is parametrized as $K(E)=K_0 (E/\GeV)^\delta$, while $b(E,\vec{r})$ is the rate of energy loss, which is dominated by the inverse Compton scattering of the positrons on the interstellar radiation field and by the synchrotron losses on the Galactic magnetic field. Information on the parameters $K_0$, $\delta$ and $L$ can be inferred from the flux measurements of other cosmic ray species, mainly from the boron-to-carbon ratio~\cite{Maurin:2001sj}. There exist, however, degeneracies among them which cannot be broken with present experiments, and therefore there is a continuous family of parameters leading to the same B/C ratio. We will choose for our analysis the MED propagation model proposed in Ref.~\cite{Donato:2003xg} from an analysis of the primary antiproton flux from dark matter annihilations, and which is characterized by $K_0=0.0112 \kpc^2/\Myr$, $\delta=0.70$ and $L=4 \kpc$ (other approaches to calculate the fluxes of cosmic positrons can be found in Refs. \cite{Moskalenko:1997gh,galprop,Evoli:2008dv,Gaggero:2013rya}). Useful parametrizations to calculate the number density of positrons using this formalism were given in Ref. \cite{Cirelli:2010xx}. Finally, the interstellar (IS) positron flux at the position of the Sun is given by:
\begin{equation}
\Phi^{\rm{DM, IS}}_{e^+}(E) = \frac{c}{4 \pi} f_{e^+}(E)\;,
\label{eq:DM-flux}
\end{equation}

The observation of the positron flux produced in dark matter annihilations or decays is hindered by the existence of still poorly known astrophysical backgrounds. The positron background contains a secondary component produced in collisions of cosmic rays in the interstellar medium.  Different calculations \cite{Moskalenko:1997gh,Delahaye:2008ua} suggest that the spectrum of secondary positrons at energies larger than 2 GeV follows a simple power law 
\begin{equation}
\Phi^{\rm sec, IS}_{e^+}(E)=C_{e^+} E^{-\gamma_{e^+}}\;,
\label{eq:sec}
\end{equation}
where $\gamma_{e^+}$ is in the range $3.3-3.7$. In addition, there may be a primary positron component produced {\it e.g.} by the interactions of high-energy photons in the strong magnetic fields of pulsars~\cite{Sturrock:1971zc,Atoian:1995ux,Chi:1995id,Grimani:2004qm,Hooper:2008kg} (other possible sources of cosmic positrons are a nearby gamma-ray burst~\cite{Ioka:2008cv} or  hadronic interactions inside the same sources that accelerate galactic cosmic rays~\cite{Blasi:2009hv}). We will assume that this potential contribution to the positron flux can be parametrized as a simple power law multiplied by an exponential cut-off:
\begin{equation}
\Phi^{\rm source, IS}_{e^+}(E)=C_s E^{-\gamma_s} \exp(-E/E_s)\;.
\label{eq:source}
\end{equation}
The total interstellar positron flux can then be expressed as:
\begin{equation}
\Phi^{\rm IS}_{e^+}(E)=\Phi^{\rm sec, IS}_{e^+}(E)+\Phi^{\rm source, IS}_{e^+}(E)+\Phi^{\rm{DM, IS}}_{e^+}(E)\;.
\end{equation}

The positron flux at low energies is significantly altered by solar modulation effects. Under the force field approximation~\cite{1967ApJ149L115G,1968ApJ1541011G}, the flux at the top of the atmosphere (TOA) is related to the flux at the heliospheric boundary by the following relation~\cite{1987AA184119P}:
\begin{equation}
\Phi^{\rm TOA}_{e^+}(E)=
\frac{E^2}{(E +\phi_{e^+})^2}
\Phi_{e^+}^{\rm IS}(E +\phi_{e^+})\;,
\label{eq:solar-modulation}
\end{equation}
where $\phi_{e^+}$ is a parameter that varies between $\sim 500\MV- 1.3 \GV$ over the eleven-year solar cycle.

\begin{figure}[t]
\begin{center}
\includegraphics[height=3.5cm]{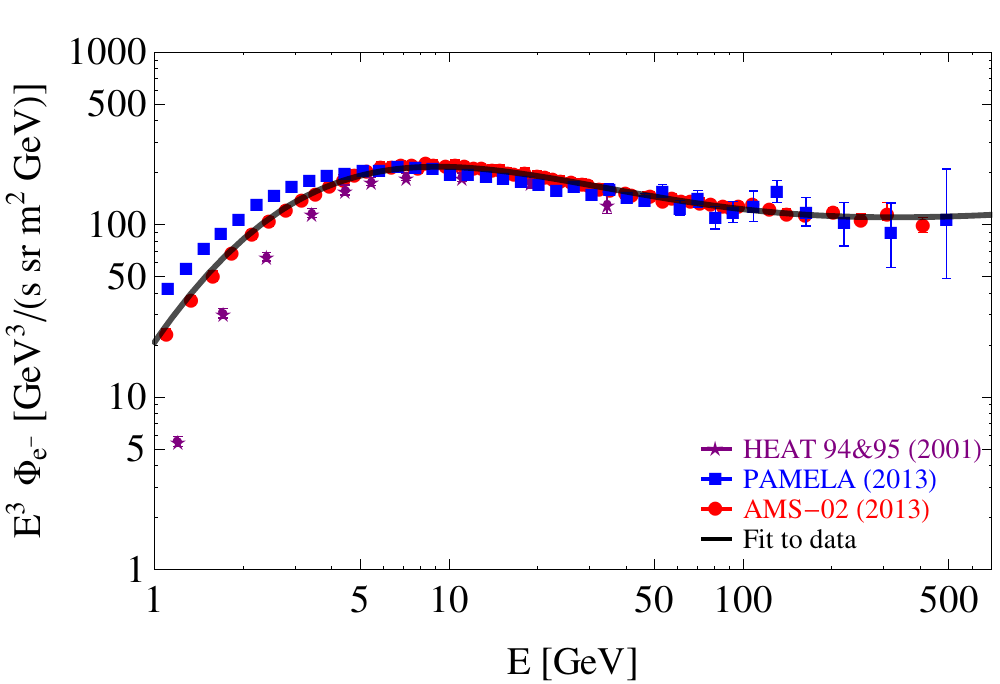}~~
\includegraphics[height=3.5cm]{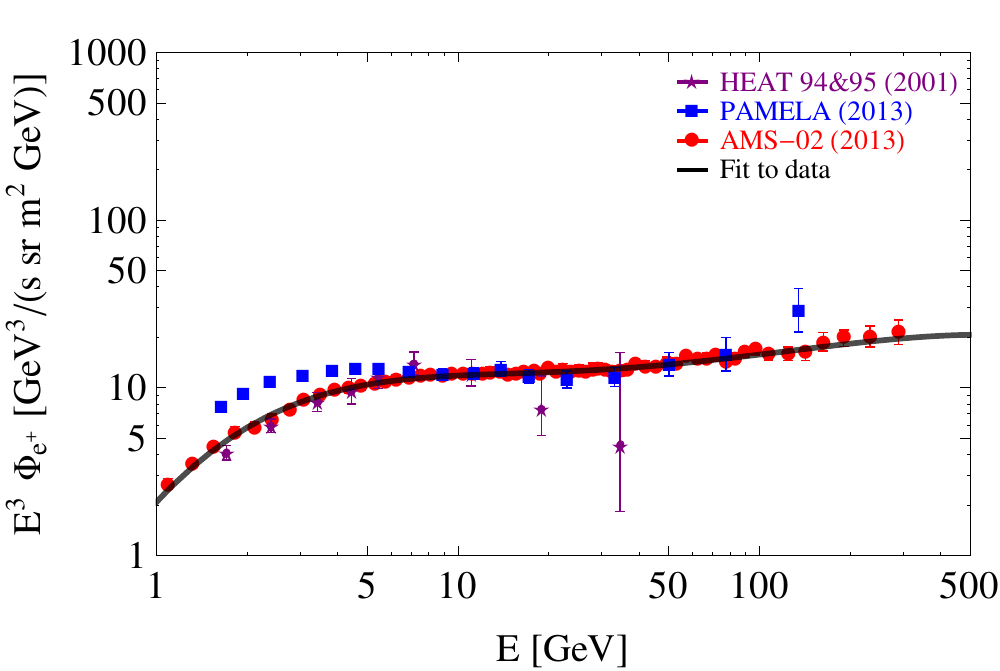}~~
\includegraphics[height=3.5cm]{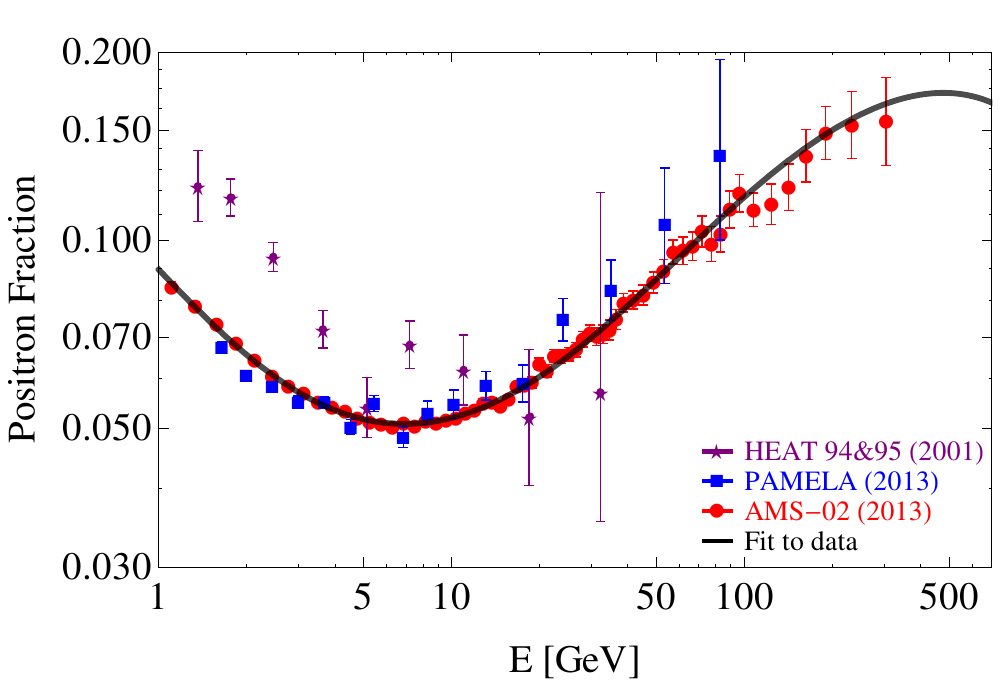}
\caption{Measurements by the HEAT, PAMELA and AMS-02 experiments of the electron flux (left plot), positron flux (middle plot) and positron fraction (right plot) together with the fit to the data of the ``background models'' described in the text.}
\label{fig:fits}
\end{center}
\end{figure}

Measurements of the positron flux are complemented by measurements of the positron fraction, defined as the positron flux over the total electron plus positron flux. From the theoretical perspective, the derivation of limits on the dark matter properties from the positron fraction requires knowledge of the electron flux, which is difficult to model. Therefore, for our analysis of the AMS-02 data on the positron fraction, we will simply fix the electron flux at the top of the atmosphere to the data measured by the AMS-02 Collaboration itself~\cite{Schael}, and which were collected in the same period as the positron data.\footnote{Note that the primary electron flux produced in the dark matter annihilations and decays is, in most models, identical to the primary positron flux, given by Eq.~(\ref{eq:DM-flux}), while the measured electron flux is about one order of magnitude larger than the positron flux. Therefore, at a given energy, we do not expect a significant contribution to the total electron flux from dark matter annihilations or decays for the parameters that saturate the observational limits on an exotic component in the positron flux.} For electron energies between 2 and 350 GeV, which is the range of interest for calculating the positron fraction, we find that the electron flux is well described by the sum of two power laws which are distorted due to solar modulation effects:
\begin{equation}
\Phi^{\rm TOA}_{e^-}(E)=  \frac{E^2}{(E+\phi_{e^-})^2}\Big[ C_1 (E+\phi_{e^-})^{-\gamma_1}+ C_2 (E+\phi_{e^-})^{-\gamma_2}\Big]\;,
\label{eq:electron-flux}
\end{equation}
where $C_1= 2.6\times 10^3 \s^{-1} \sr^{-1} \m^{-2} \GeV^{-1}$, $\gamma_1 = 3.83$,  $C_2=35\s^{-1} \sr^{-1} \m^{-2} \GeV^{-1}$, $\gamma_2= 2.83$ and $\phi_{e^-}= 1.3 \GV$ (see Fig. \ref{fig:fits}, left plot).

\section{Limits on the dark matter annihilation cross section and lifetime}
\label{sec:limits}

\begin{figure}[t]
\begin{center}
\includegraphics[height=5.2cm]{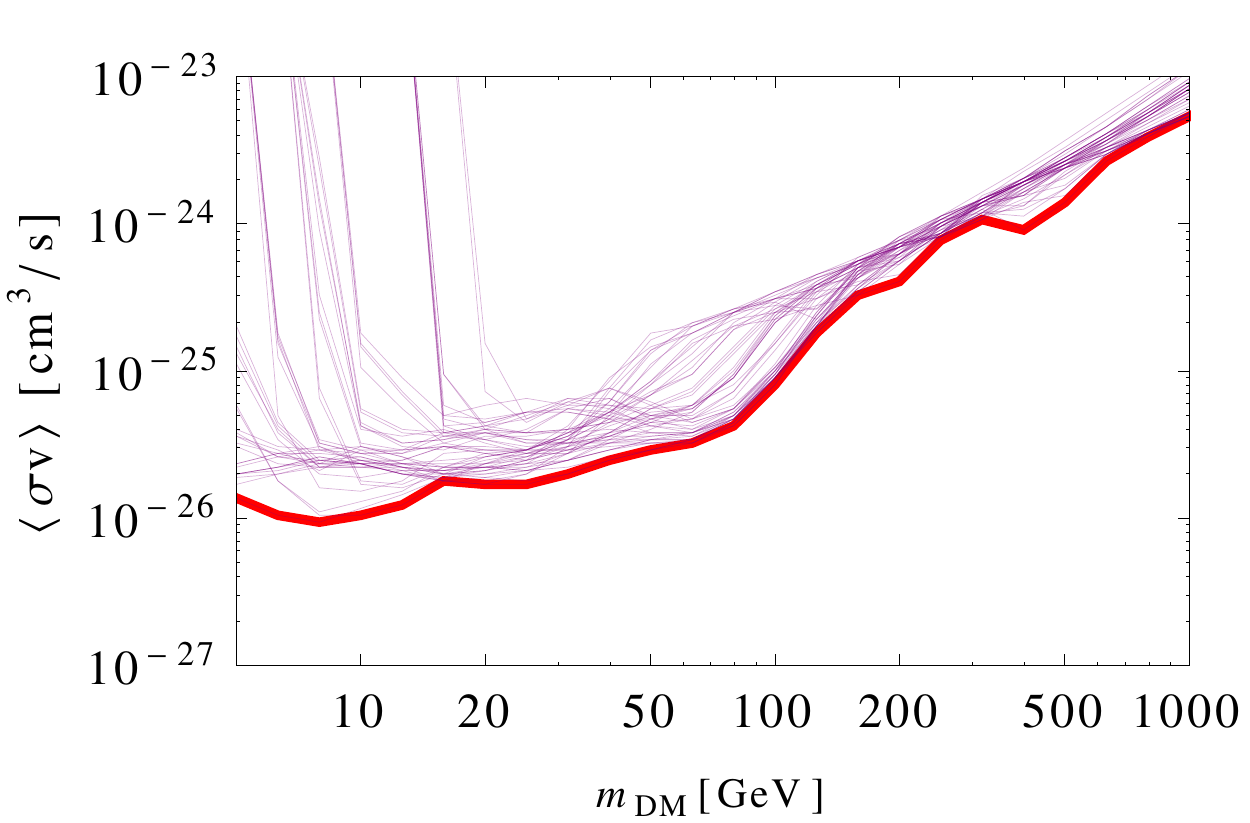}
\includegraphics[height=5.2cm]{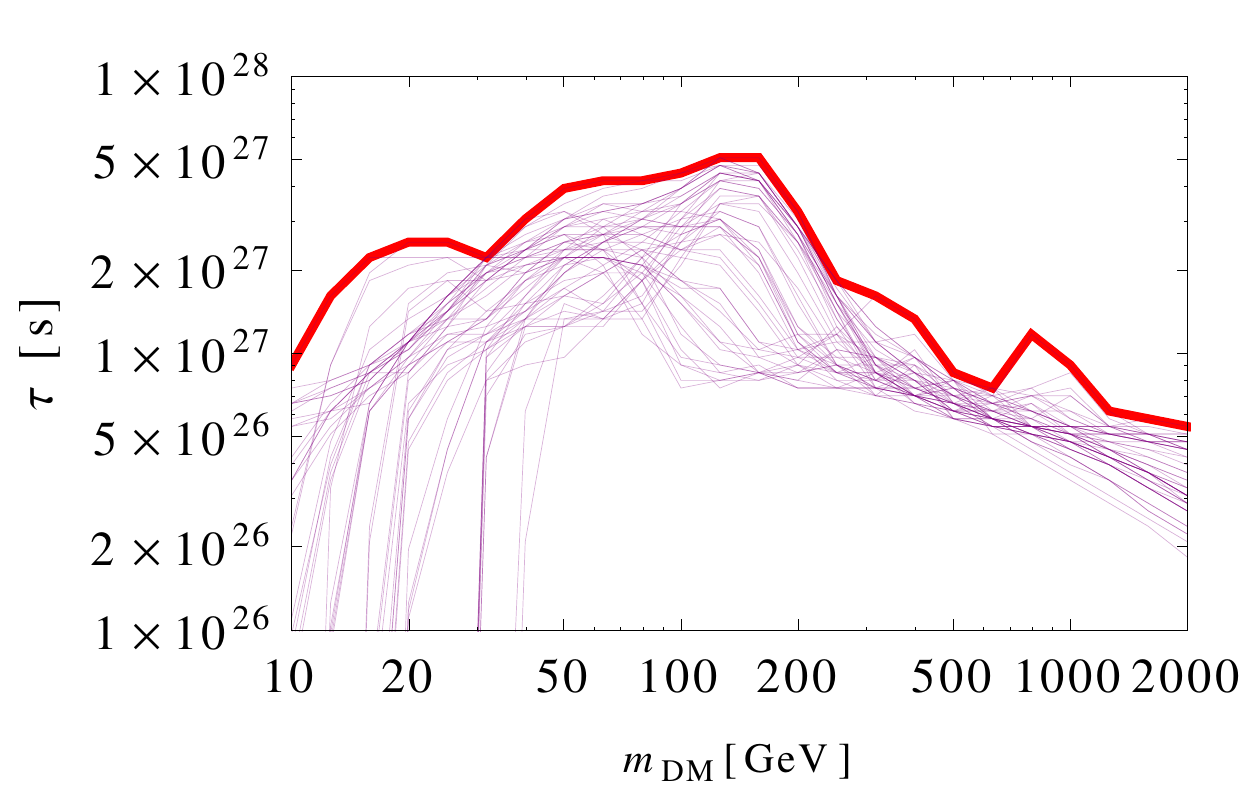}
\caption{2$\sigma$ limits on the dark matter annihilation cross section (left plot) and lifetime (right plot) derived from the AMS-02 data on the positron flux, assuming the final state $\mu^+\mu^-$. The thin purple lines show the limits for each of the 53 energy windows used in our analysis, while the thick red line shows the best limit for each dark matter mass.}
\label{fig:scans}
\end{center}
\end{figure}

To set limits on the dark matter annihilation cross section and lifetime, we first use the measurements of the positron flux from the AMS-02~\cite{Schael} Collaboration. We will make the well-motivated assumption that the background receives a contribution from secondary positrons produced in cosmic ray spallations with a spectrum given by Eq.~(\ref{eq:sec}) and possibly a contribution from astrophysical sources with a spectrum given by  Eq.~(\ref{eq:source}). The form of these spectra is well justified when $E\gtrsim 2\GeV$, hence we will use only data satisfying this condition in our analysis. We then select an energy window in the positron data and perform a $\chi^2$ test with the six parameters appearing in the flux at the top of the atmosphere, $C_{e^+}$, $\gamma_{e^+}$, $C_s$, $\gamma_s$, $E_s$ and $\phi_{e^+}$ with the restrictions $C_{e^+},~C_s \geq 0$, $3.7> \gamma_{e^+}> 3.3$, $\gamma_{e^+}>\gamma_s$, $E_s>0$, $1.3\GV> \phi_{e^+} >0.5 \GV$. For windows reaching energies larger than $\sim 10\GeV$ the secondary contribution does not provide by itself a good fit to the data, and hence it is necessary to include a source contribution, Eq.~(\ref{eq:source}), with a harder spectrum than the secondary contribution. Specifically, for the AMS-02 measurements of the positron flux ~\cite{Schael}, we obtain the best fit for $E=2-350\GeV$  when $C_{e^+}=72 \s^{-1} \sr^{-1} \m^{-2} \GeV^{-1}$, $\gamma_{e^+}=3.70$, $C_s=1.6 \s^{-1} \sr^{-1} \m^{-2} \GeV^{-1}$, $\gamma_s=2.51$, $E_s=1.0\TeV$ and $\phi_{e^+}=0.93\GV$ (see Fig. \ref{fig:fits}, middle plot).\footnote{Note that the best fit for the solar modulation parameter of our background models is different for electrons and for positrons. Charge-sign-dependent solar modulation effects were also noted from the PAMELA data \cite{Gast-Schael,Maccione:2012cu}.} We then introduce a contribution to the positron flux from the annihilations or decays of dark matter particles with mass $m_{\rm DM}$ and annihilation cross section $\langle \sigma v \rangle$ or lifetime $\tau$ and recalculate the $\chi^2$ of the best fit model. The $2\sigma$ exclusion line for this specific energy window is obtained by determining, for a given final state and dark matter mass, the cross section and lifetime which gives a $\chi^2$ which exceeds the $\chi^2$ of the ``pure background'' model by more than 4. Lastly, we select, for a given dark matter mass, the most stringent limit among those obtained from sampling various energy windows. Our procedure is illustrated in Fig.~\ref{fig:scans} for the case of annihilations and decays into a $\mu^+\mu^-$ pair from our analysis of the AMS-02 measurements of the positron flux. The plot shows the limits on the annihilation cross section (left plot) and lifetime (right plot) obtained from sampling 53 energy windows (shown as thin purple lines), together with the curve that gives the most stringent limit for each mass (shown as a thick red line). 

\begin{figure}[t]
\begin{center}
\includegraphics[height=5.2cm]{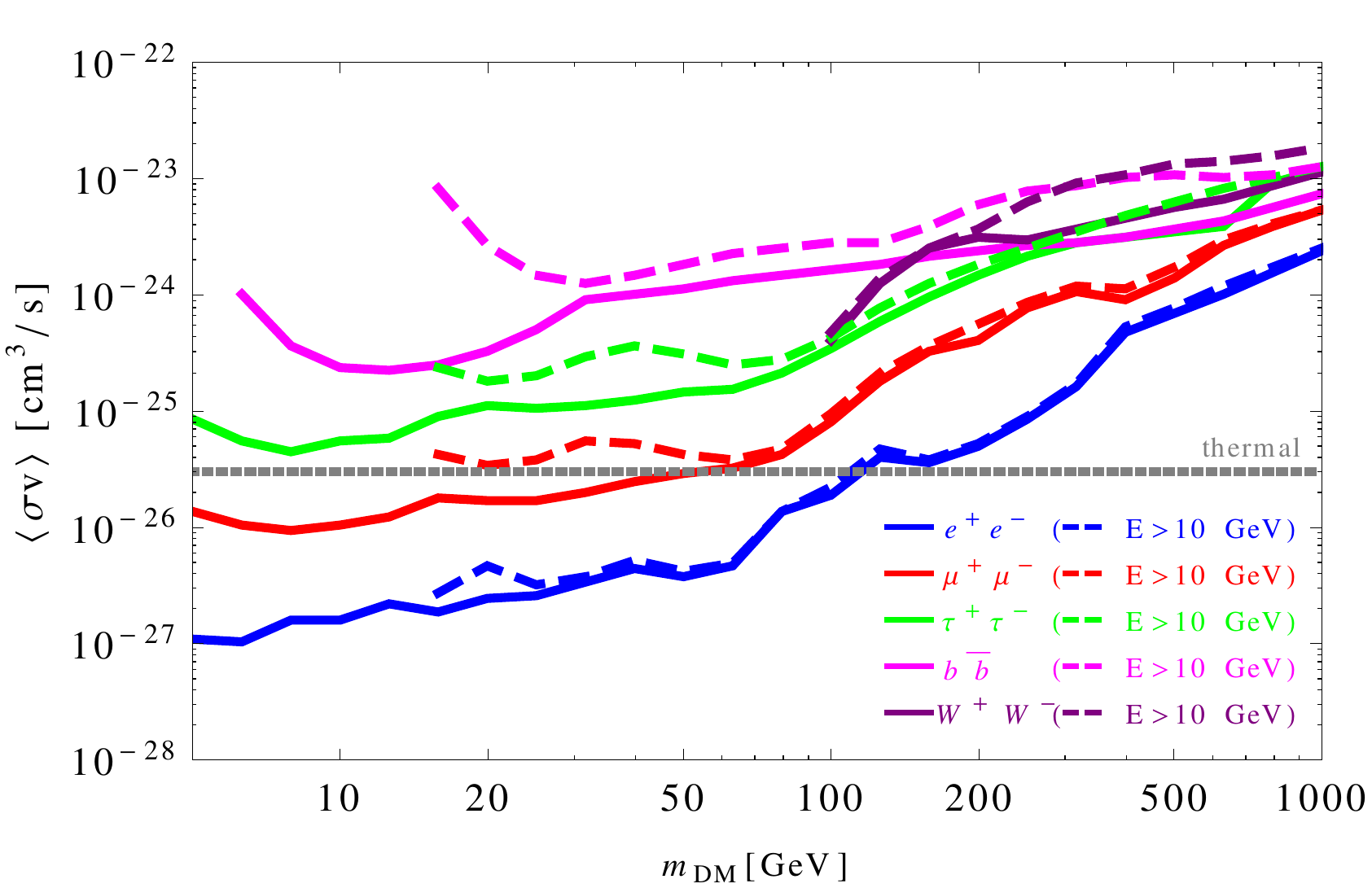}
\includegraphics[height=5.2cm]{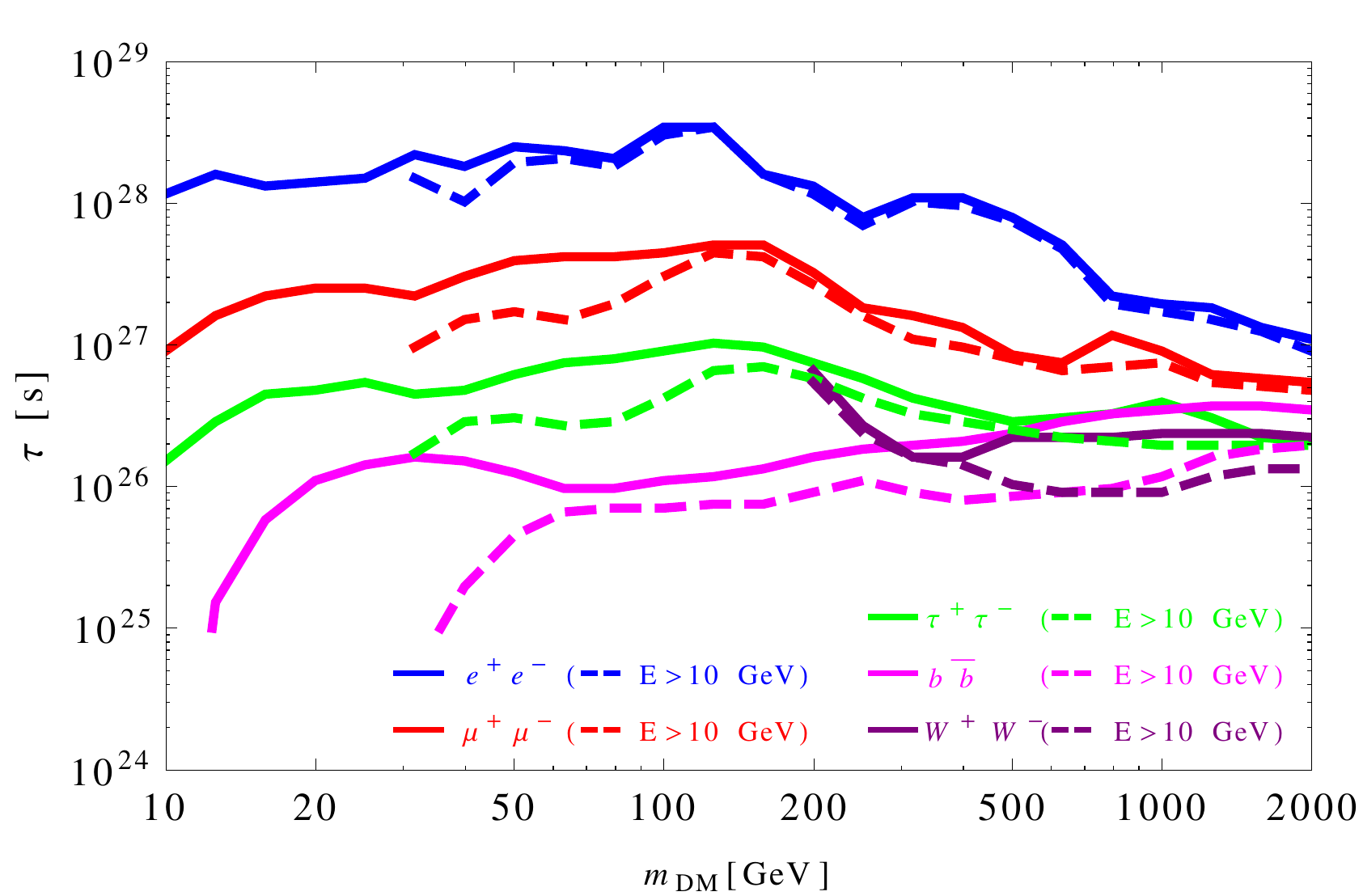}
\caption{Limits on the annihilation cross section (left plot) and the lifetime (right plot) derived from the AMS-02 data on the positron flux, assuming the final states $e^+e^-$, $\mu^+\mu^-$, $\tau^+\tau^-$, $b\bar b$ and $W^+W^-$ and the MED propagation model. The limits shown as solid lines were derived from sampling over various energy windows, while the dashed lines are from considering those windows including only data with energies above 10 GeV.}
\label{fig:limits-flux-allchannels}
\end{center}
\end{figure}

\begin{figure}[t]
\begin{center}
\includegraphics[height=5.2cm]{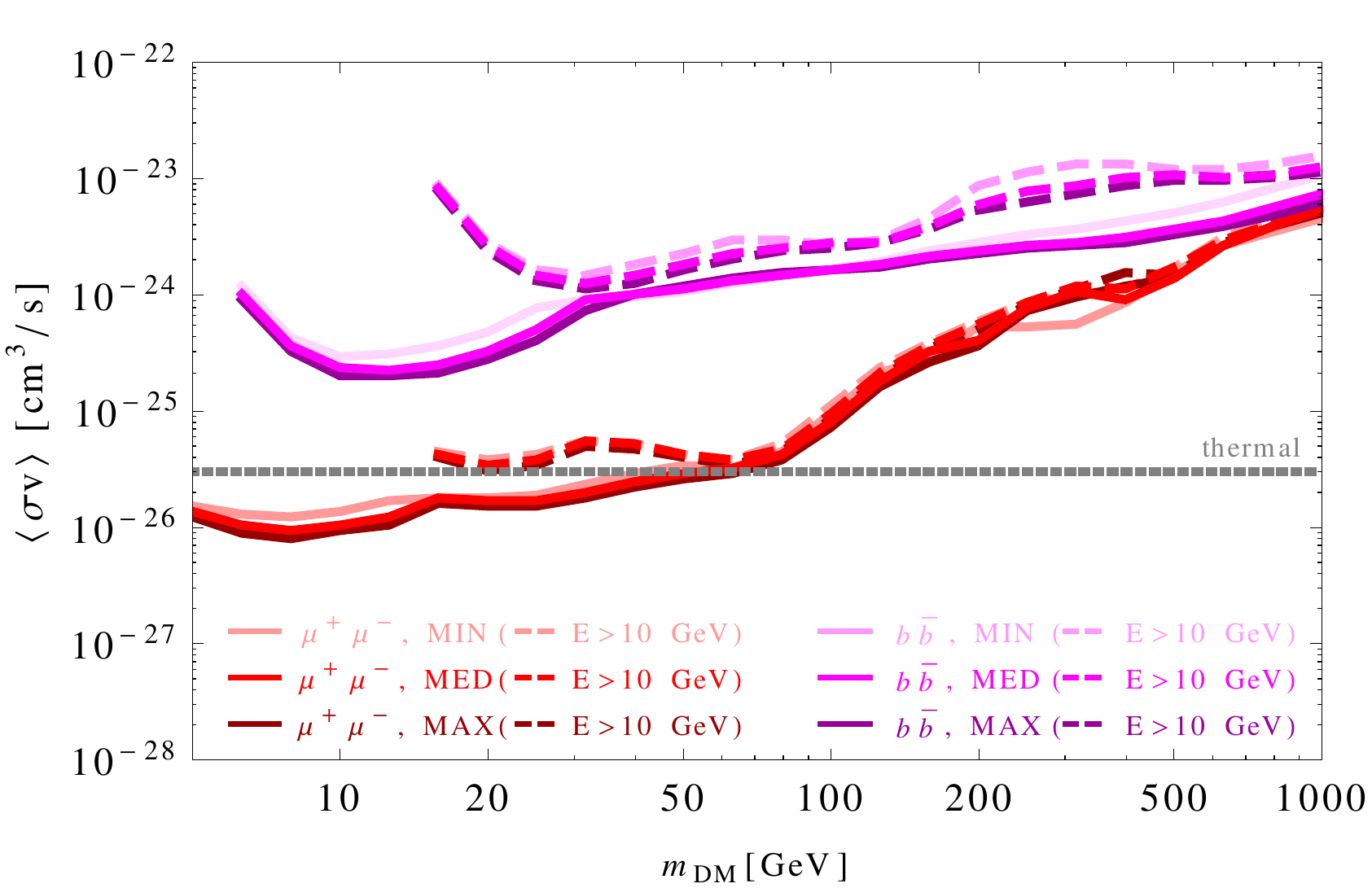}~~
\includegraphics[height=5.2cm]{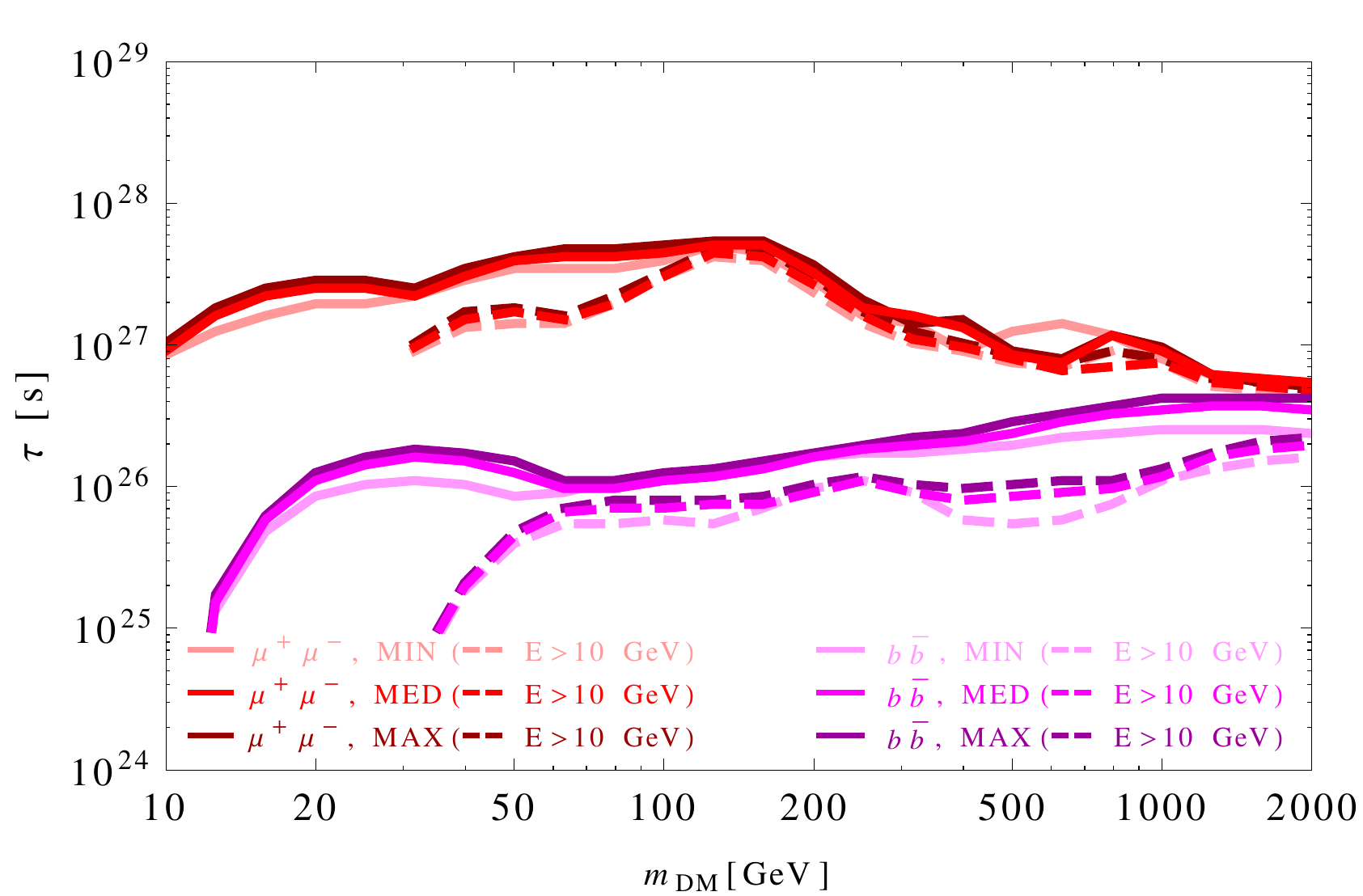}
\caption{Same as Fig.~\ref{fig:limits-flux-allchannels}, showing the impact of the choice of the propagation model on the limits for the final states $\mu^+\mu^-$ and $b\bar b$.}
\label{fig:prop-uncertainty}
\end{center}
\end{figure}

Following this procedure, we have calculated the limits on the annihilation cross section and lifetime for the final states $e^+ e^-$, $\mu^+\mu^-$, $\tau^+\tau^-$, $b \bar b$ and $W^+W^-$; the results are presented in Fig.~\ref{fig:limits-flux-allchannels}. The limits shown with solid lines correspond to the best limits sampling over all windows, while those shown with dashed lines were derived selecting windows containing only energies larger than 10 GeV. The latter limits are mildly affected by the modeling of the solar modulation and are therefore more robust. The limits are very strong for the $e^+e^-$ final state and, in fact, present measurements of the positron flux probe cross sections smaller than the ``thermal cross section'', $\langle \sigma v \rangle=3\times 10^{-26}\cm^3/\s$, for dark matter masses below 100 GeV; in the case of the  $\mu^+\mu^-$ final state the limits are below the thermal cross section for $m_{\rm DM}\lesssim 60\GeV$. We have calculated the limits also for the Navarro-Frenk-White profile and the isothermal profile, as well as for the MIN and MAX propagation models presented in Ref.~\cite{Donato:2003xg}. We have found that the conclusions are mildly affected by the choice of propagation parameters, as shown in Fig.\ref{fig:prop-uncertainty}, and by the choice of the halo profile.

\begin{figure}[t]
\begin{center}
\includegraphics[height=5.2cm]{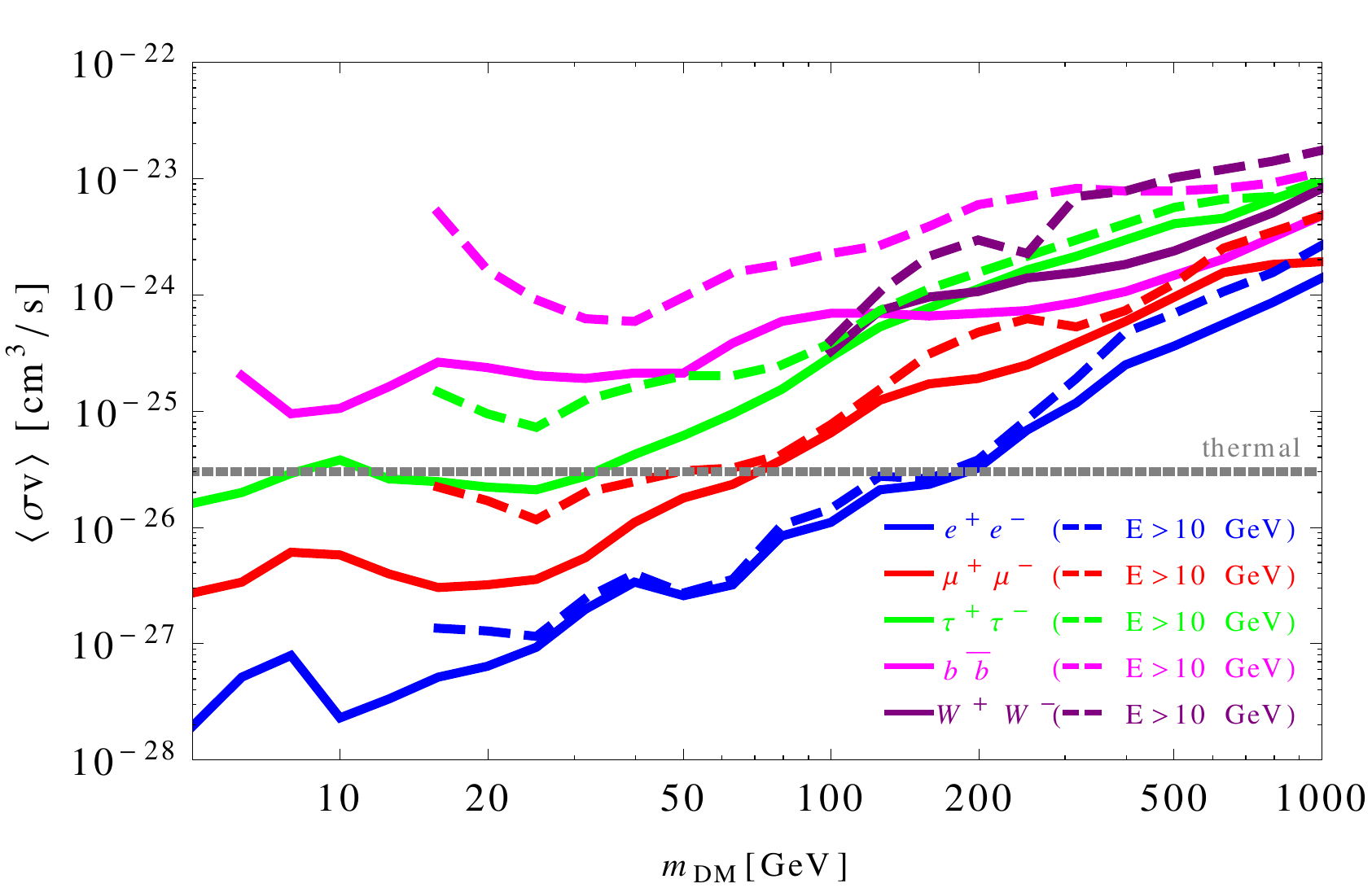}
\includegraphics[height=5.2cm]{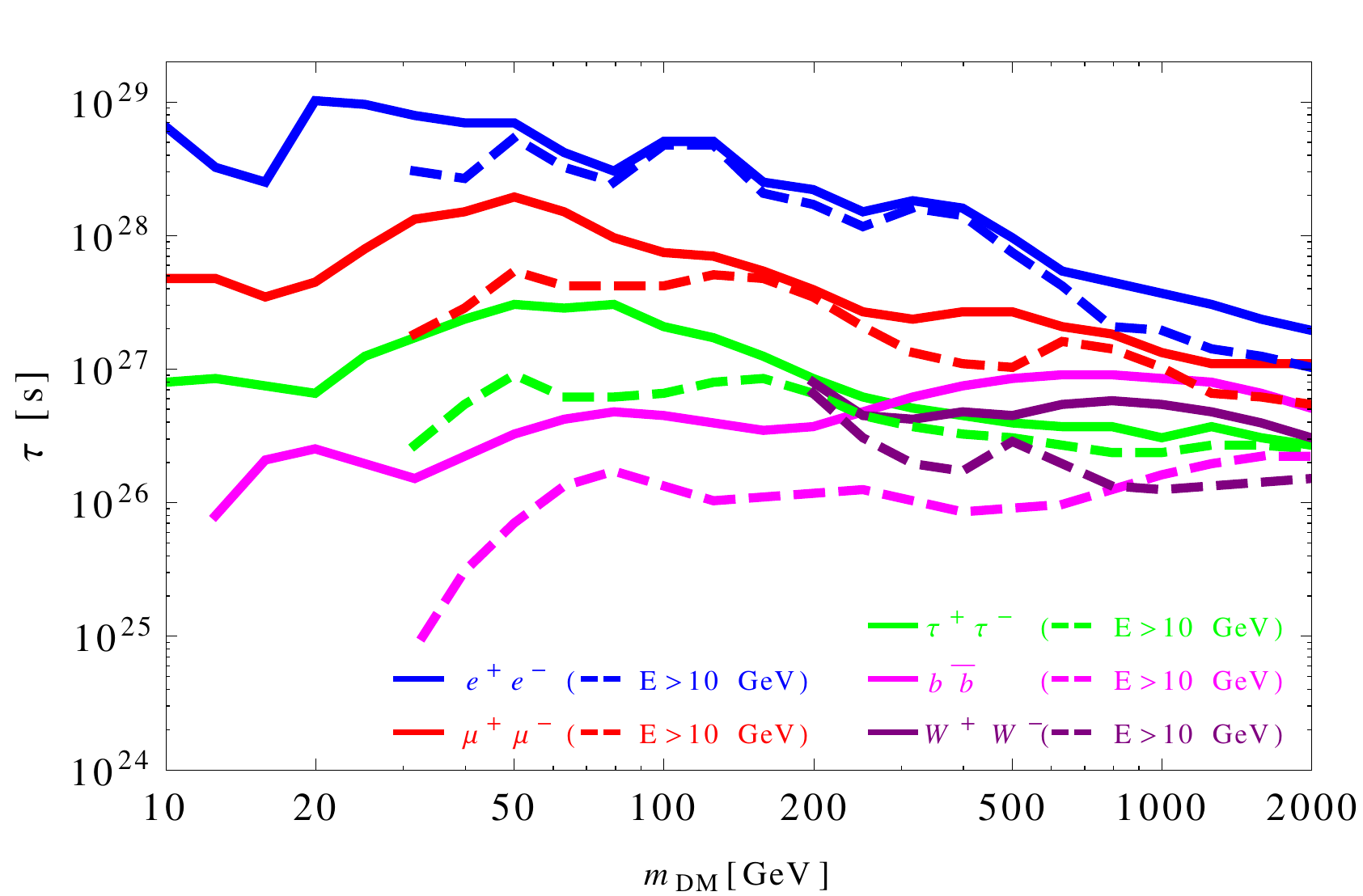}
\caption{Same as Fig. \ref{fig:limits-flux-allchannels} but using the AMS-02 data on the positron fraction.}
\label{fig:PF-allchannels}
\end{center}
\end{figure}

Following the same procedure, we have derived for comparison limits on the annihilation cross section and the lifetime using data on the positron fraction published by AMS-02. To calculate the positron fraction, we fix the electron flux to the parametrization given in Eq.~(\ref{eq:electron-flux}) derived from the AMS-02 measurements~\cite{Schael}. We now postulate a background positron flux consisting of a power law plus a second power law with a cut-off, modified using the force field approximation to take into account solar modulation effects. This model reproduces very well the AMS-02 data in the range $E=2 \GeV-350 \GeV$ when  $C_{e^+}=61 \s^{-1} \sr^{-1} \m^{-2} \GeV^{-1}$, $\gamma_{e^+}=3.70$, $C_s=2.4 \s^{-1} \sr^{-1} \m^{-2} \GeV^{-1}$, $\gamma_s=2.60$, $E_s=1.4\TeV$ and $\phi_{e^+}=0.83$, giving a $\chi^2/{\rm d.o.f}=27.78/52$, and leaving little room for an exotic component in the positron fraction (see Fig.~\ref{fig:fits}, right plot).\footnote{Note that the parameters of the ``only-background model'' for the positron flux inferred from the AMS-02 measurements of the positron fraction are compatible with those inferred from the positron flux itself.} As before, we now introduce an additional component in the positron flux and recalculate the minimum $\chi^2$ of the new model in various energy windows. Finally, we determine the $2\sigma$ exclusion lines for each energy window and select, for each dark matter mass, the best limit. The resulting limits for the final states $e^+e^-$, $\mu^+\mu^-$, $\tau^+\tau^-$, $b\bar b$ and $W^+ W^-$ are shown in Fig. \ref{fig:PF-allchannels} for dark matter annihilations (left plot) and dark matter decays (right plot).\footnote{The limits derived in this paper from the measurements of the positron fraction are weaker than those reported in Ref.~\cite{Bergstrom:2013jra} due mostly to two factors: firstly, to the modeling of the positron energy loss, which results in fluxes smaller by approximately a factor of 2 in our analysis and hence to weaker limits, and secondly, to the different choices of the electron plus positron flux used in the calculation of the positron fraction, namely the Fermi-LAT data in the case of Ref.~\cite{Bergstrom:2013jra} and the AMS-02 data in our case. The larger $e^-+e^+$ flux measured by AMS-02 compared to Fermi-LAT \cite{Bertucci} again results in weaker limits.}

\begin{figure}[t]
\begin{center}
\includegraphics[height=5.2cm]{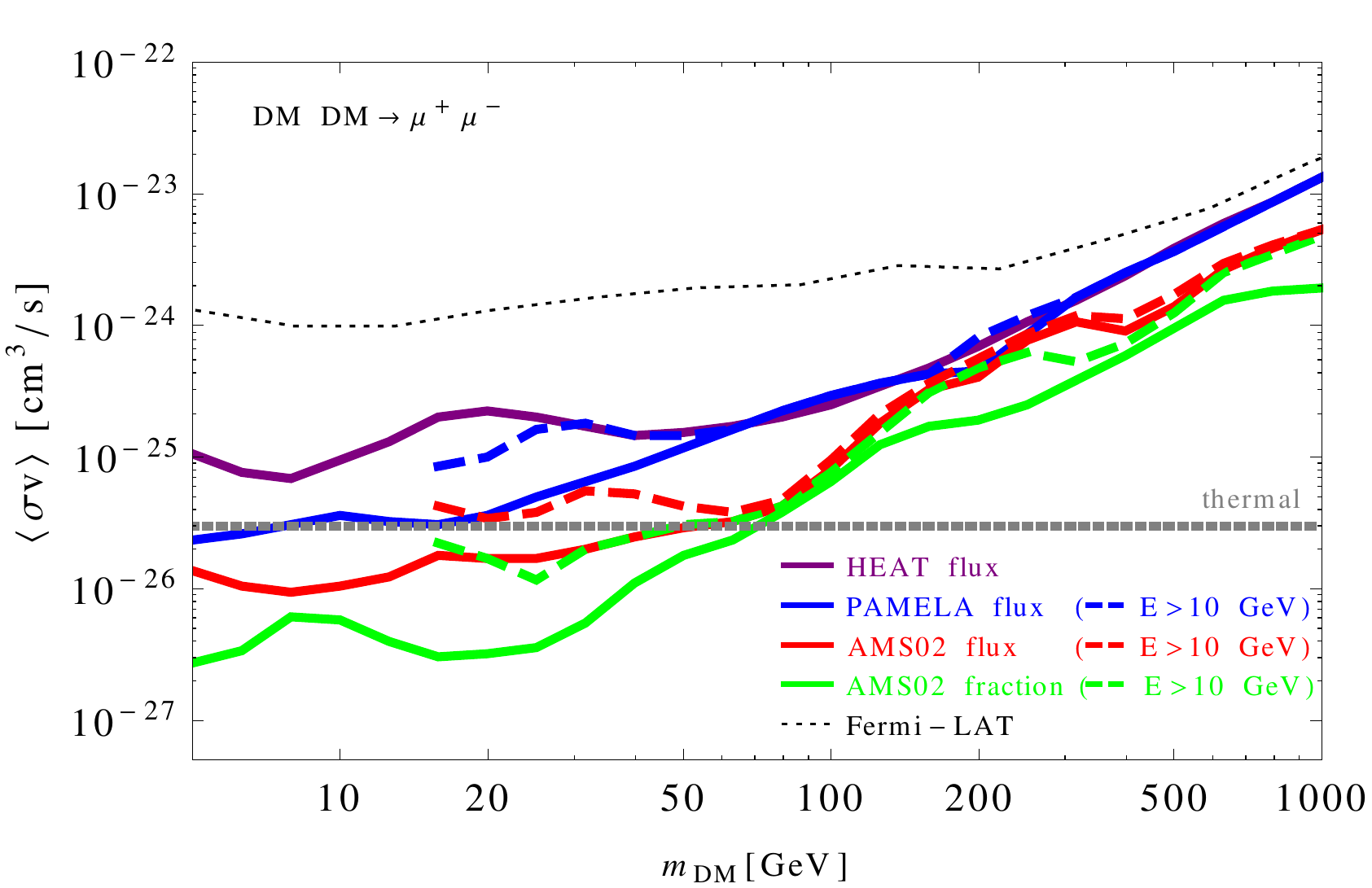}
\includegraphics[height=5.2cm]{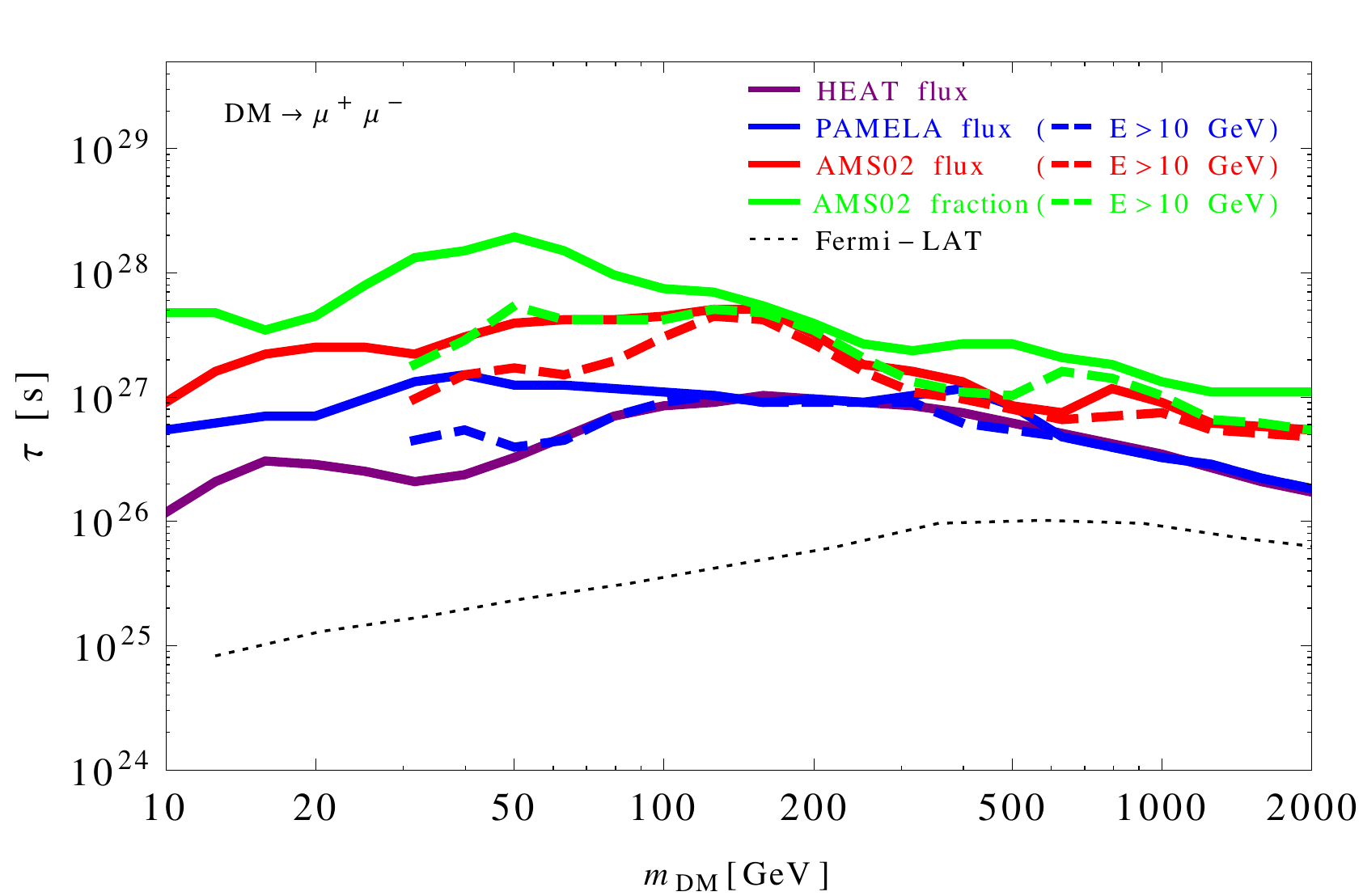}\\
\includegraphics[height=5.2cm]{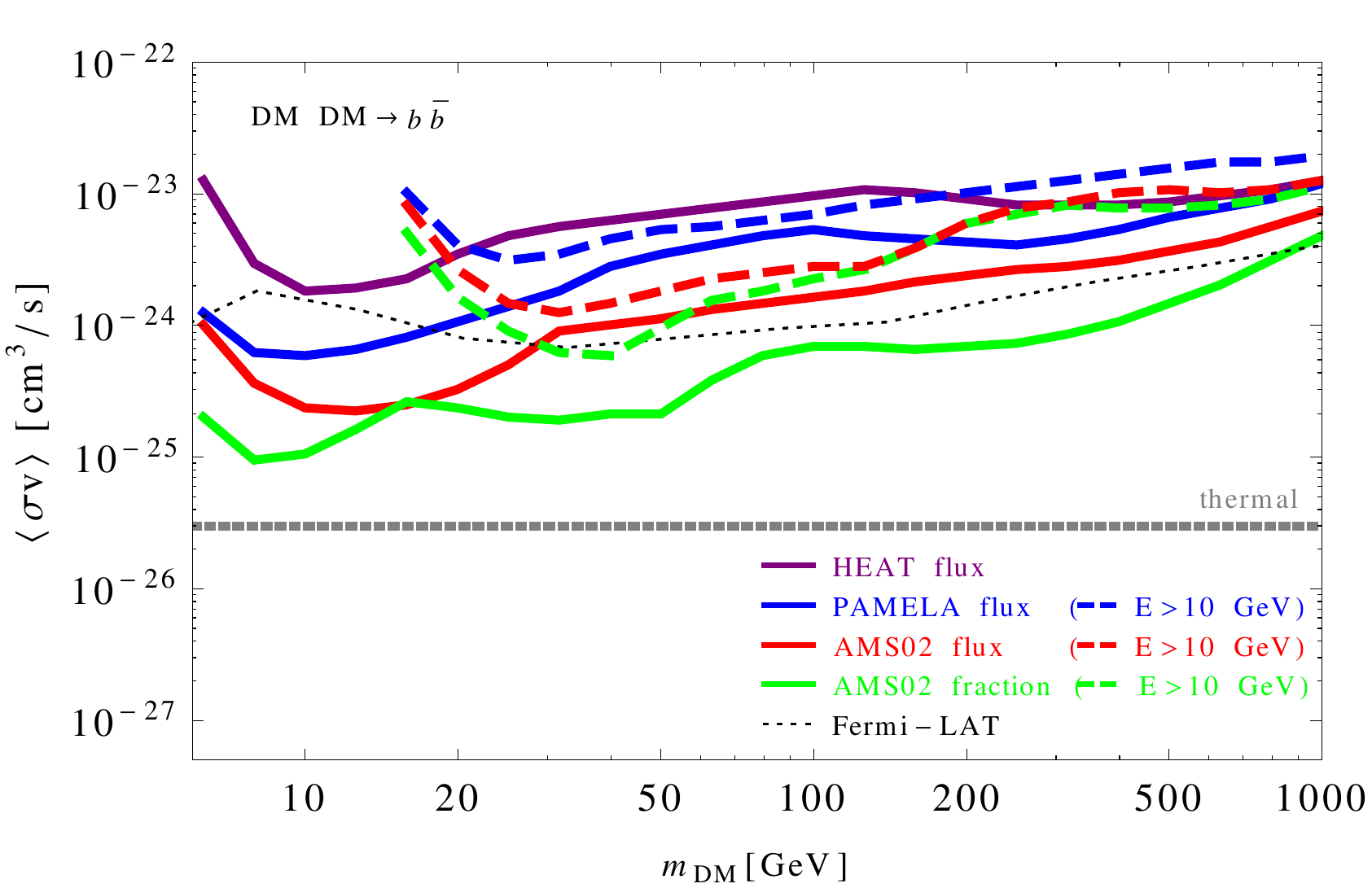}
\includegraphics[height=5.2cm]{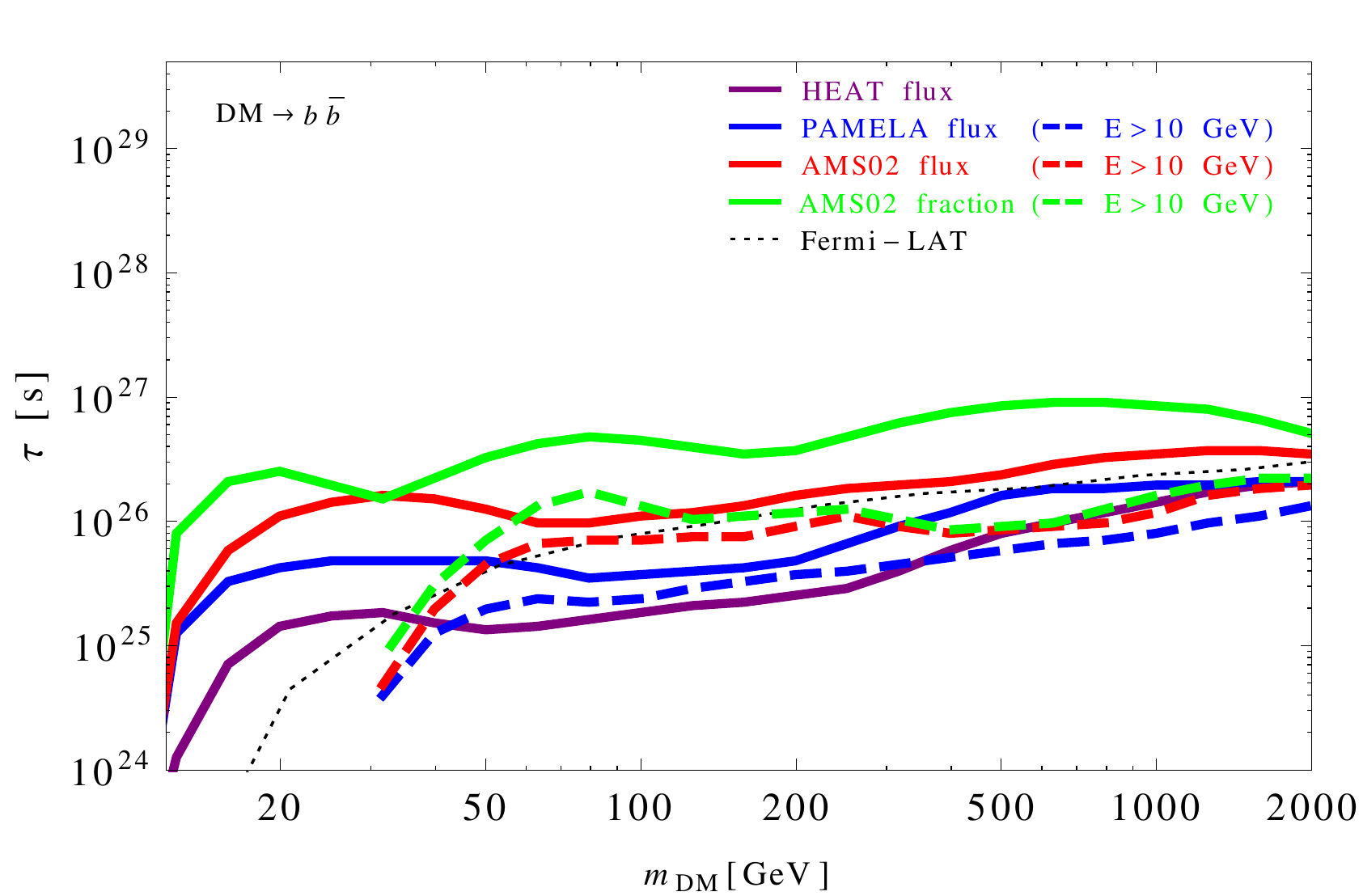}\\
\caption{Limits on the annihilation cross section (left plot) and the lifetime (right plot) derived in this paper from the HEAT, PAMELA and AMS-02 measurements of the positron flux and the AMS-02 measurements of the positron fraction, assuming the final state $\mu^+\mu^-$ (upper plots) and  $b\bar b$ (lower plots), together with the limits on the same parameters from gamma-ray observations.}
\label{fig:limits-muons-different experiments}
\end{center}
\end{figure}

Lastly, we compare in Fig.~\ref{fig:limits-muons-different experiments} the limits on the annihilation cross section or lifetime derived from the positron flux measured by HEAT~\cite{DuVernois:2001bb}, PAMELA~\cite{Adriani:2013uda} and AMS-02~\cite{Schael}, as well as from the positron fraction measured by AMS-02~\cite{Aguilar:2013qda}, assuming the final states $\mu^+\mu^-$  (upper plots) and  $b\bar b$ (lower plots). We also show the limits on the annihilation cross section and lifetime reported in Ref.\cite{Ackermann:2012qk} from measurements of the diffuse gamma-ray background  (which are strongly correlated to the positron limits since both the gamma-ray and the positron fluxes have the same source term), as well as the limits on the cross section reported in Ref.\cite{Ackermann:2013yva} from a combined analysis of 25 Milky Way satellite galaxies and the limits on the lifetime in Ref.\cite{Cirelli:2012ut} from the extragalactic gamma-ray background. For annihilations into $\mu^+\mu^-$, the limits on the cross section derived from the AMS-02 measurements of the positron flux are, at $m_{\rm DM}=10\GeV$, two orders  of magnitude stronger than those reported by the Fermi-LAT Collaboration from measurements of the diffuse gamma-ray background (and one order of magnitude in the case of dwarf galaxies), while, at $m_{\rm DM}=1\TeV$, about a factor of 3 stronger. On the other hand, the limits derived from the measurements of the positron fraction are stronger than the limits from the positron flux by a factor ranging between 1 and 4, depending on the mass. The same conclusions apply to the limits on the dark matter lifetime. It is remarkable that the HEAT measurements already provide better limits in this annihilation channel than the Fermi-LAT data. Annihilations and decays into $b\bar b$ generate softer positrons and harder gamma rays than $\mu^+\mu^-$, and therefore we expect weaker limits on the annihilation cross section and lifetime from positron measurements and stronger limits from gamma-ray measurements. Indeed, and as shown in the figure, for this channel the Fermi-LAT limits derived from observations of the diffuse background are comparable to those derived from the AMS-02 measurements of the positron flux, and slightly worse than those derived from the positron fraction, while the limits derived from dwarf galaxy observations are stronger.

\section{Conclusions}
\label{sec:conclusions}
We have derived limits on the dark matter annihilation cross section and lifetime using, for the first time, measurements of the positron flux, as well as using measurements of the positron fraction. In deriving the limits, we have made the well-motivated assumption that a background positron flux exists. We have parametrized the interstellar background flux above 2 GeV as a sum of two components: a first one with an energy spectrum following a simple power law, as suggested by theoretical calculations of the secondary positron flux, and a second one following a simple power law times an exponential cut-off, as generically expected from astrophysical positron sources. Finally, the cosmic positron flux at the top of the atmosphere was calculated employing the force field approximation. The parameters entering in the background positron flux have then been fitted to the data together with the primary flux from dark matter annihilations or decays. We have further optimized the search by applying this procedure to 53 energy windows of different sizes and then choosing for each dark matter mass the best among all 53 limits. The resulting limits are very strong and are competitive to those derived from the positron fraction and also competitive (and for some channels stronger) than the limits reported by the Fermi-LAT Collaboration on those same final states from gamma-ray flux measurements. Specifically, we find that the measurements of the positron flux by AMS-02 probe the thermal cross section when the dark matter particles annihilate only into $e^+ e^-$ or into into $\mu^+\mu^-$,  provided the dark matter mass is smaller than 100 GeV  or 60 GeV, respectively.

\vspace{0.5cm}
\section*{Acknowledgements}
 This work was partially supported by the DFG cluster of excellence ``Origin and Structure of the Universe'' and by the Graduiertenkolleg ``Particle Physics at the Energy Frontier of New Phenomena.'' The research of J.S. has been supported at IAP by  the ERC project  267117 (DARK) hosted by Universit\'e Pierre et Marie Curie - Paris 6   and at JHU by NSF grant OIA-1124403. A.I. and J.S. thank the Kavli Institute for Theoretical Physics in Santa Barbara, California, for hospitality during the initial stages of this work, and to Miguel Pato for very useful discussions.



\bibliographystyle{JHEP-mod}
\bibliography{positron}

\providecommand{\href}[2]{#2}\begingroup\raggedright\begin{thebibliography}{10}

\bibitem{Boezio:2000zz}
M.~Boezio, P.~Carlson, T.~Francke, N.~Weber, M.~Suffert, {\em et.~al.}, {\it
  {The Cosmic-Ray Electron and Positron Spectra Measured at 1 AU during Solar
  Minimum Activity}},  {\em Astrophys.J.} {\bf 532} (2000) 653--669.

\bibitem{DuVernois:2001bb}
M.~DuVernois, S.~Barwick, J.~Beatty, A.~Bhattacharyya, C.~Bower, {\em et.~al.},
  {\it {Cosmic ray electrons and positrons from 1-GeV to 100-GeV: Measurements
  with HEAT and their interpretation}},  {\em Astrophys.J.} {\bf 559} (2001)
  296--303.

\bibitem{Aguilar:2007yf}
{\bf AMS-01 Collaboration}, M.~Aguilar {\em et.~al.}, {\it {Cosmic-ray positron
  fraction measurement from 1 to 30-GeV with AMS-01}},  {\em Phys.Lett.} {\bf
  B646} (2007) 145--154, [\href{http://xxx.lanl.gov/abs/astro-ph/0703154}{{\tt
  astro-ph/0703154}}].

\bibitem{Adriani:2008zr}
{\bf PAMELA Collaboration}, O.~Adriani {\em et.~al.}, {\it {An anomalous
  positron abundance in cosmic rays with energies 1.5-100 GeV}},  {\em Nature}
  {\bf 458} (2009) 607--609, [\href{http://xxx.lanl.gov/abs/0810.4995}{{\tt
  arXiv:0810.4995}}].

\bibitem{Adriani:2013uda}
{\bf PAMELA Collaboration}, O.~Adriani {\em et.~al.}, {\it {The cosmic-ray
  positron energy spectrum measured by PAMELA}},
  \href{http://xxx.lanl.gov/abs/1308.0133}{{\tt arXiv:1308.0133}}.

\bibitem{Aguilar:2013qda}
{\bf AMS Collaboration}, M.~Aguilar {\em et.~al.}, {\it {First Result from the
  Alpha Magnetic Spectrometer on the International Space Station: Precision
  Measurement of the Positron Fraction in Primary Cosmic Rays of 0.5-350 GeV}},
   {\em Phys.Rev.Lett.} {\bf 110} (2013), no.~14 141102.

\bibitem{Schael}
S. Schael for the AMS Collaboration, \textit{Precision measurements of the
  electron spectrum and the positron spectrum with AMS}. Proceedings of the
  33rd International Cosmic Ray Conference, Rio de Janeiro, 2-9 July 2013.

\bibitem{Cirelli:2008pk}
M.~Cirelli, M.~Kadastik, M.~Raidal, and A.~Strumia, {\it {Model-independent
  implications of the e+-, anti-proton cosmic ray spectra on properties of Dark
  Matter}},  {\em Nucl.Phys.} {\bf B813} (2009) 1--21,
  [\href{http://xxx.lanl.gov/abs/0809.2409}{{\tt arXiv:0809.2409}}].

\bibitem{Ibarra:2008jk}
A.~Ibarra and D.~Tran, {\it {Decaying Dark Matter and the PAMELA Anomaly}},
  {\em JCAP} {\bf 0902} (2009) 021,
  [\href{http://xxx.lanl.gov/abs/0811.1555}{{\tt arXiv:0811.1555}}].

\bibitem{Meade:2009iu}
P.~Meade, M.~Papucci, A.~Strumia, and T.~Volansky, {\it {Dark Matter
  Interpretations of the e+- Excesses after FERMI}},  {\em Nucl.Phys.} {\bf
  B831} (2010) 178--203, [\href{http://xxx.lanl.gov/abs/0905.0480}{{\tt
  arXiv:0905.0480}}].

\bibitem{Kopp:2013eka}
J.~Kopp, {\it {Constraints on dark matter annihilation from AMS-02 results}},
  \href{http://xxx.lanl.gov/abs/1304.1184}{{\tt arXiv:1304.1184}}.

\bibitem{Feng:2013zca}
L.~Feng, R.-Z. Yang, H.-N. He, T.-K. Dong, Y.-Z. Fan, {\em et.~al.}, {\it
  {AMS-02 positron excess: new bounds on dark matter models and hint for
  primary electron spectrum hardening}},
  \href{http://xxx.lanl.gov/abs/1303.0530}{{\tt arXiv:1303.0530}}.

\bibitem{Yuan:2013eja}
Q.~Yuan, X.-J. Bi, G.-M. Chen, Y.-Q. Guo, S.-J. Lin, {\em et.~al.}, {\it
  {Implications of the AMS-02 positron fraction in cosmic rays}},
  \href{http://xxx.lanl.gov/abs/1304.1482}{{\tt arXiv:1304.1482}}.

\bibitem{Cholis:2013psa}
I.~Cholis and D.~Hooper, {\it {Dark Matter and Pulsar Origins of the Rising
  Cosmic Ray Positron Fraction in Light of New Data From AMS}},
  \href{http://xxx.lanl.gov/abs/1304.1840}{{\tt arXiv:1304.1840}}.

\bibitem{Jin:2013nta}
H.-B. Jin, Y.-L. Wu, and Y.-F. Zhou, {\it {Implications of the first AMS-02
  measurement for dark matter annihilation and decay}},
  \href{http://xxx.lanl.gov/abs/1304.1997}{{\tt arXiv:1304.1997}}.

\bibitem{Bergstrom:2013jra}
L.~Bergstrom, T.~Bringmann, I.~Cholis, D.~Hooper, and C.~Weniger, {\it {New
  limits on dark matter annihilation from AMS cosmic ray positron data}},  {\em
  Phys.Rev.Lett.} (2013) [\href{http://xxx.lanl.gov/abs/1306.3983}{{\tt
  arXiv:1306.3983}}].

\bibitem{Hooper:2012gq}
D.~Hooper and W.~Xue, {\it {Possibility of Testing the Light Dark Matter
  Hypothesis with the Alpha Magnetic Spectrometer}},  {\em Phys.Rev.Lett.} {\bf
  110} (2013) 041302, [\href{http://xxx.lanl.gov/abs/1210.1220}{{\tt
  arXiv:1210.1220}}].

\bibitem{Salati:2010rc}
P.~Salati, F.~Donato, and N.~Fornengo, {\it {Indirect Dark Matter Detection
  with Cosmic Antimatter}},  \href{http://xxx.lanl.gov/abs/1003.4124}{{\tt
  arXiv:1003.4124}}.

\bibitem{Ibarra:2013cra}
A.~Ibarra, D.~Tran, and C.~Weniger, {\it {Indirect Searches for Decaying Dark
  Matter}},  \href{http://xxx.lanl.gov/abs/1307.6434}{{\tt arXiv:1307.6434}}.

\bibitem{Sjostrand:2007gs}
T.~Sjostrand, S.~Mrenna, and P.~Z. Skands, {\it {A Brief Introduction to PYTHIA
  8.1}},  {\em Comput.Phys.Commun.} {\bf 178} (2008) 852--867,
  [\href{http://xxx.lanl.gov/abs/0710.3820}{{\tt arXiv:0710.3820}}].

\bibitem{Navarro:2003ew}
J.~F. Navarro, E.~Hayashi, C.~Power, A.~Jenkins, C.~S. Frenk, {\em et.~al.},
  {\it {The Inner structure of Lambda-CDM halos 3: Universality and asymptotic
  slopes}},  {\em Mon.Not.Roy.Astron.Soc.} {\bf 349} (2004) 1039,
  [\href{http://xxx.lanl.gov/abs/astro-ph/0311231}{{\tt astro-ph/0311231}}].

\bibitem{Graham:2005xx}
A.~W. Graham, D.~Merritt, B.~Moore, J.~Diemand, and B.~Terzic, {\it {Empirical
  models for Dark Matter Halos. I. Nonparametric Construction of Density
  Profiles and Comparison with Parametric Models}},  {\em Astron.J.} {\bf 132}
  (2006) 2685--2700, [\href{http://xxx.lanl.gov/abs/astro-ph/0509417}{{\tt
  astro-ph/0509417}}].

\bibitem{Navarro:2008kc}
J.~F. Navarro, A.~Ludlow, V.~Springel, J.~Wang, M.~Vogelsberger, {\em et.~al.},
  {\it {The Diversity and Similarity of Cold Dark Matter Halos}},
  \href{http://xxx.lanl.gov/abs/0810.1522}{{\tt arXiv:0810.1522}}.

\bibitem{Catena:2009mf}
R.~Catena and P.~Ullio, {\it {A novel determination of the local dark matter
  density}},  {\em JCAP} {\bf 1008} (2010) 004,
  [\href{http://xxx.lanl.gov/abs/0907.0018}{{\tt arXiv:0907.0018}}].

\bibitem{Weber:2009pt}
M.~Weber and W.~de~Boer, {\it {Determination of the Local Dark Matter Density
  in our Galaxy}},  {\em Astron.Astrophys.} {\bf 509} (2010) A25,
  [\href{http://xxx.lanl.gov/abs/0910.4272}{{\tt arXiv:0910.4272}}].

\bibitem{Salucci:2010qr}
P.~Salucci, F.~Nesti, G.~Gentile, and C.~Martins, {\it {The dark matter density
  at the Sun's location}},  {\em Astron.Astrophys.} {\bf 523} (2010) A83,
  [\href{http://xxx.lanl.gov/abs/1003.3101}{{\tt arXiv:1003.3101}}].

\bibitem{Pato:2010yq}
M.~Pato, O.~Agertz, G.~Bertone, B.~Moore, and R.~Teyssier, {\it {Systematic
  uncertainties in the determination of the local dark matter density}},  {\em
  Phys.Rev.} {\bf D82} (2010) 023531,
  [\href{http://xxx.lanl.gov/abs/1006.1322}{{\tt arXiv:1006.1322}}].

\bibitem{Iocco:2011jz}
F.~Iocco, M.~Pato, G.~Bertone, and P.~Jetzer, {\it {Dark Matter distribution in
  the Milky Way: microlensing and dynamical constraints}},  {\em JCAP} {\bf
  1111} (2011) 029, [\href{http://xxx.lanl.gov/abs/1107.5810}{{\tt
  arXiv:1107.5810}}].

\bibitem{Berezinskii}
V. S. Berezinskii, S. V. Buolanov, V. A. Dogiel, V. L. Ginzburg, V. S. Ptuskin,
  \textit{Astrophysics of Cosmic Rays} (North-Holland, 1990).

\bibitem{Maurin:2001sj}
D.~Maurin, F.~Donato, R.~Taillet, and P.~Salati, {\it {Cosmic rays below z=30
  in a diffusion model: new constraints on propagation parameters}},  {\em
  Astrophys.J.} {\bf 555} (2001) 585--596,
  [\href{http://xxx.lanl.gov/abs/astro-ph/0101231}{{\tt astro-ph/0101231}}].

\bibitem{Donato:2003xg}
F.~Donato, N.~Fornengo, D.~Maurin, and P.~Salati, {\it {Antiprotons in cosmic
  rays from neutralino annihilation}},  {\em Phys.Rev.} {\bf D69} (2004)
  063501, [\href{http://xxx.lanl.gov/abs/astro-ph/0306207}{{\tt
  astro-ph/0306207}}].

\bibitem{Moskalenko:1997gh}
I.~Moskalenko and A.~Strong, {\it {Production and propagation of cosmic ray
  positrons and electrons}},  {\em Astrophys.J.} {\bf 493} (1998) 694--707,
  [\href{http://xxx.lanl.gov/abs/astro-ph/9710124}{{\tt astro-ph/9710124}}].

\bibitem{galprop}
http://galprop.stanford.edu/.

\bibitem{Evoli:2008dv}
C.~Evoli, D.~Gaggero, D.~Grasso, and L.~Maccione, {\it {Cosmic-Ray Nuclei,
  Antiprotons and Gamma-rays in the Galaxy: a New Diffusion Model}},  {\em
  JCAP} {\bf 0810} (2008) 018, [\href{http://xxx.lanl.gov/abs/0807.4730}{{\tt
  arXiv:0807.4730}}].

\bibitem{Gaggero:2013rya}
D.~Gaggero, L.~Maccione, G.~Di~Bernardo, C.~Evoli, and D.~Grasso, {\it
  {Three-Dimensional Model of Cosmic-Ray Lepton Propagation Reproduces Data
  from the Alpha Magnetic Spectrometer on the International Space Station}},
  {\em Phys.Rev.Lett.} {\bf 111} (2013) 021102,
  [\href{http://xxx.lanl.gov/abs/1304.6718}{{\tt arXiv:1304.6718}}].

\bibitem{Cirelli:2010xx}
M.~Cirelli, G.~Corcella, A.~Hektor, G.~Hutsi, M.~Kadastik, {\em et.~al.}, {\it
  {PPPC 4 DM ID: A Poor Particle Physicist Cookbook for Dark Matter Indirect
  Detection}},  {\em JCAP} {\bf 1103} (2011) 051,
  [\href{http://xxx.lanl.gov/abs/1012.4515}{{\tt arXiv:1012.4515}}].

\bibitem{Delahaye:2008ua}
T.~Delahaye, F.~Donato, N.~Fornengo, J.~Lavalle, R.~Lineros, {\em et.~al.},
  {\it {Galactic secondary positron flux at the Earth}},  {\em
  Astron.Astrophys.} {\bf 501} (2009) 821--833,
  [\href{http://xxx.lanl.gov/abs/0809.5268}{{\tt arXiv:0809.5268}}].

\bibitem{Sturrock:1971zc}
P.~Sturrock, {\it {A Model of pulsars}},  {\em Astrophys.J.} {\bf 164} (1971)
  529.

\bibitem{Atoian:1995ux}
A.~Atoian, F.~Aharonian, and H.~Volk, {\it {Electrons and positrons in the
  galactic cosmic rays}},  {\em Phys.Rev.} {\bf D52} (1995) 3265--3275.

\bibitem{Chi:1995id}
X.~Chi, E.~Young, and K.~Cheng, {\it {Pulsar-wind origin of cosmic ray
  positrons}},  {\em Astrophys.J.} {\bf 459} (1995) L83--L86.

\bibitem{Grimani:2004qm}
C.~Grimani, {\it {Pulsar birthrate set by cosmic-ray positron observations}},
  {\em Astron.Astrophys.} {\bf 418} (2004) 649--653.

\bibitem{Hooper:2008kg}
D.~Hooper, P.~Blasi, and P.~D. Serpico, {\it {Pulsars as the Sources of High
  Energy Cosmic Ray Positrons}},  {\em JCAP} {\bf 0901} (2009) 025,
  [\href{http://xxx.lanl.gov/abs/0810.1527}{{\tt arXiv:0810.1527}}].

\bibitem{Ioka:2008cv}
K.~Ioka, {\it {A Gamma-Ray Burst/Pulsar for Cosmic-Ray Positrons with a Dark
  Matter-like Spectrum}},  {\em Prog.Theor.Phys.} {\bf 123} (2010) 743--755,
  [\href{http://xxx.lanl.gov/abs/0812.4851}{{\tt arXiv:0812.4851}}].

\bibitem{Blasi:2009hv}
P.~Blasi, {\it {The origin of the positron excess in cosmic rays}},  {\em
  Phys.Rev.Lett.} {\bf 103} (2009) 051104,
  [\href{http://xxx.lanl.gov/abs/0903.2794}{{\tt arXiv:0903.2794}}].

\bibitem{1967ApJ149L115G}
L.~J. Gleeson and W.~I. Axford, {\it {Cosmic Rays in the Interplanetary
  Medium}},  {\em Astrophys.J.} {\bf 149} (1967) L115.

\bibitem{1968ApJ1541011G}
L.~J. Gleeson and W.~I. Axford, {\it {Solar Modulation of Galactic Cosmic
  Rays}},  {\em Astrophys.J.} {\bf 154} (1968) 1011.

\bibitem{1987AA184119P}
J.~S. Perko, {\it {Solar modulation of galactic antiprotons}},  {\em Astron.
  Astrophys.} {\bf 184} (1987) 119--121.

\bibitem{Gast-Schael}
H. Gast, S. Schael, \textit{Charge-dependent solar modulation in light of the
  recent PAMELA data}. Talk presented at the 31st International Cosmic Ray
  Conference, Lodz, 7-15 July 2009.

\bibitem{Maccione:2012cu}
L.~Maccione, {\it {Low energy cosmic ray positron fraction explained by
  charge-sign dependent solar modulation}},  {\em Phys.Rev.Lett.} {\bf 110}
  (2013), no.~8 081101, [\href{http://xxx.lanl.gov/abs/1211.6905}{{\tt
  arXiv:1211.6905}}].

\bibitem{Bertucci}
B. Bertucci for the AMS Collaboration, \textit{The $e^-+e^+$ flux measurements
  with the AMS experiment on ISS}. Proceedings of the 33rd International Cosmic
  Ray Conference, Rio de Janeiro, 2-9 July 2013.

\bibitem{Ackermann:2012qk}
{\bf Fermi-LAT Collaboration}, M.~Ackermann {\em et.~al.}, {\it {Fermi LAT
  Search for Dark Matter in Gamma-ray Lines and the Inclusive Photon
  Spectrum}},  {\em Phys.Rev.} {\bf D86} (2012) 022002,
  [\href{http://xxx.lanl.gov/abs/1205.2739}{{\tt arXiv:1205.2739}}].

\bibitem{Ackermann:2013yva}
{\bf Fermi-LAT Collaboration}, M.~Ackermann {\em et.~al.}, {\it {Dark Matter
  Constraints from Observations of 25 Milky Way Satellite Galaxies with the
  Fermi Large Area Telescope}},  \href{http://xxx.lanl.gov/abs/1310.0828}{{\tt
  arXiv:1310.0828}}.

\bibitem{Cirelli:2012ut}
M.~Cirelli, E.~Moulin, P.~Panci, P.~D. Serpico, and A.~Viana, {\it {Gamma ray
  constraints on Decaying Dark Matter}},  {\em Phys.Rev.} {\bf D86} (2012)
  083506, [\href{http://xxx.lanl.gov/abs/1205.5283}{{\tt arXiv:1205.5283}}].

\end{thebibliography}\endgroup

\end{document}